\documentclass[aps,prl,twocolumn,groupedaddress,showpacs,floatfix]{revtex4-1}

\usepackage{graphicx}
\usepackage{hyperref}

\begin{document}
\def\be{\begin{equation}}
\def\ee{\end{equation}}
\def\ba{\begin{eqnarray}}
\def\ea{\end{eqnarray}}
\def\parderiv#1#2{\frac{\partial #1}{\partial #2}}

\title{Falling atoms}

\author{Neil Ashby}
\affiliation{University of Colorado, Boulder, CO 80309}
\affiliation{National Institute of Standards and Technology, Boulder, CO 80305}
\email{ashby@boulder.nist.gov}

\begin{abstract}
	Atomic fountain clocks launch laser-cooled balls of atoms upwards to some height $h$ in the local gravitational field where they experience both second-order Doppler shifts and gravitational frequency shifts.  It is shown in this paper that the net shift, relative to a reference at the launch point, is $ g h/3 c^2$.   We derive the next most significant correction to this expression, and show that the value of $g$ should be corrected for Coriolis and centripetal effects.  
\end{abstract}


\maketitle

\section{Introduction}

	Atoms in an atomic fountain are launched upwards through a cavity that prepares their state, rise upward to some height $h$ above the cavity, and then fall back down through the cavity and are then detected.  While in flight their gravitational potentials and velocities change, so they suffer variations in gravitational frequency shift and second-order Doppler shift.  We shall take a simplified view of this process in the next section.  Assuming the gravitational field is static and uniform and that the motion of the atoms is entirely vertical, we shall show that the average fractional frequency shift is $ g h/(3 c^2)$ plus a small correction. Thus for the purpose of estimating the corrections due to second-order Doppler and gravitational frequency shift, the atom can be considered to be at one-third the fountain height above a similar but stationary reference atom.  

	On the other hand consider the process from the point of view of an Earth-Centered Inertial (ECI) frame.  Atoms in flight are actually revolve about the earth in an almost-Keplerian orbit with high eccentricity, and are perturbed by earth's multipole gravitational potentials. A reference atom that coincides with the orbiting atom at the beginning and end of the flight can be considered to be at rest at a reference position on earth's surface and are carried along with earth's rotation.  In this picture the orbiting atom has a high, nearly horizontal velocity so the second-order Doppler shift contribution is quite different.  The principle of equivalence implies that the same frequency shift, relative to the reference atom, will occur no matter which of these points of view is adopted.  However it is at once evident that the gravitational field strength near earth's surface is not exactly uniform, and that in an earth fixed rotating frame of reference (ECEF) the orbiting atoms will be subject to centripetal and coriolis forces.  The leading contributions due to these effects are calculated in Sections 3-6.

\section{Fountain in a static uniform gravitational field}

	If we assume the gravitational field can be described as uniform and static, with numerical value $g$ (in m/s$^2$), then the atoms are basically in free fall.  If the initial launch velocity at the reference point is $v_0$, then their vertical velocity and position as functions of time are
\ba
v=v_0-gt\,;\\
z=v_0 t-\frac{1}{2}g t^2\,.
\ea
The apex of the trajectory is reached at time $t=v_0/g$, and at this time the height is
\be
h=\frac{v_0^2}{2g}\,.
\ee

{\it Second-order Doppler shift\/}.  The fractional frequency shift due to time dilation, or ``second-order Doppler shift," is 
\be
\frac{\Delta f}{f}_{{\rm Doppler}}=-\frac{1}{2}\frac{v^2}{c^2}=-\frac{1}{2}\frac{(v_0-g t)^2}{c^2}\,.
\ee
We perform an average of this shift over the time of flight, which we denote by
\be
\frac{\Delta f}{f}_{{\rm Doppler}}=-\big<\frac{1}{2}\frac{(v_0-g t)^2}{c^2}\big>\,.
\ee
The average of a power of the time over some time interval is given by
\be\label{eq6}
<t^n>=\frac{1}{t_f}\int_0^{t_f} t^n\,dt=\frac{1}{n+1}t_f^n\,.
\ee
Here $t_f=2 v_0/g$.  The result of a straightforward computation gives the average fractional second-order Doppler shift as
\be
\frac{\Delta f}{f}\bigg|_{{\rm Doppler}}=-\frac{1}{3}\frac{g h}{c^2}\,.
\ee

{\it Gravitational frequency shift\/}.  The fractional frequency shift of an oscillator at height $z$ above a reference in a static uniform gravitational field of strength $g$ is $g z/c^2$.  The frequency shift during the time of flight will be
\ba
\frac{\Delta  f}{f}\bigg|_{{\rm Gravitation}}=\big<\frac{g}{c^2}(v_0 t-\frac{1}{2}g t^2)  \big>\\
\hbox to 2in{}=\frac{2}{3}\frac{g h}{c^2}\,.
\ea
The combined fractional frequency shift due to these two effects is:
\be
\frac{\Delta f}{f}=\frac{1}{3}\frac{g h}{c^2}\,. 
\ee
Thus, half the gravitational contribution to the shift is cancelled by time dilation.  

\section{Transformation to Reference Frame at Earth's Surface}

In an actual atomic fountain laboratory on the surface of the rotating earth, the gravitational field is not uniform, and the direction of the local vertical does not coincide with the radius vector from Earth's center.  An atom launched upwards experiences Coriolis and Centripetal forces that depend on velocity and position.  In the present section we investigate the question: What is the next most significant correction to the fractional frequency shift due to time dilation and gravitation, beyond the simple picture discussed in the previous section?

	Earth's rotation is closely tied to Earth's oblateness and the largest higher multipole moment, the quadrupole moment, that is characterized by a quadrupole moment coefficient $ J_2$.  Keeping this term in the expansion of Earth's gravitational potential, the potential is still axially symmetric (as it is if only the even multipole moments are retained), but the next term in the expansion is smaller by a factor of about $10^3$.  In the following discussion we shall therefore consider earth's gravitational field to be axially symmetric.

	We approach this problem by writing the Lagrangian of an atom in a reference frame which is centered at the surface of the earth, at radius $R$ and geocentric colatitude $\theta$, and which is tilted by an angle $\lambda$ where $\lambda$ is the geographic latitude. The difference between $\theta$ and $\lambda$ is illustrated in Fig. 1. In this reference frame, the angle of tilt will be chosen so that the acceleration of an atom at rest at the origin is entirely vertical, including true gravitational field gradients and centripetal forces.

	We start with Earth-Centered Inertial (ECI) coordinates $(x_0, y_0, z_0)$ with $z_0$ parallel to earth's rotation axis, and first transform to Earth-Fixed, Earth-Centered (ECEF) coordinates $(x_1,y_1,z_1)$ by means of a rotation through angle $\omega t$, where $\omega$ is the angular rotational velocity of the earth.  Secondly we move the origin of coordinates to the laboratory at the surface of the earth by a translation; we call these new coordinates $(x_2,y_2,z_2)$.  Lastly we tilt the coordinates around the $y_2$ axis ($y_2$ points eastward) so that the final direction of the $z-$axis is pointing in the geographic vertical direction.  The new coordinates are denoted $(x,y,z)$; these are the coordinates that the fountain experiments are performed in; an atom at rest at $z=0$ falls in the $-z$ direction.

	In the ECI frame that we start with, the kinetic energy of a particle of mass $m_i$ is
\be
K=\frac{1}{2}m_i(\dot x_0^2 + \dot y_0^2+\dot z_0^2)\,.
\ee

	{\it Transformation to rotating coordinates\/.} The transformation to ECEF coordinates is
\ba
x_0=&x_1 \cos \omega t - y_1 \sin \omega t\,;\nonumber\\
y_0=&x_1 \sin \omega t + y_1 \cos \omega t\,;\\
z_0=&z_1\hbox to 1in{}\,. \nonumber
\ea
After differentiating the find the new velocities, the kinetic energy in the ECEF frame becomes
\ba
K=\frac{1}{2}m_i\big(\dot x_1^2 + \dot y_1^2+ \dot z_1^2 +\omega^2(x_1^2+y_1^2)\nonumber\\ +2\omega(x_1 \dot y_1-y_1 \dot x_1)\big)\,.
\ea
Thus, additional centripetal and coriolis terms appear in the kinetic energy.

	{\it Translation to Earth's surface\/.}  The coordinate system is now shifted by translation, in the rotating frame, to a point on earth's surface at distance $R$ from the center, with geocentric colatitude $\theta$.  Here $\theta$ is the usual angle measured down from the pole in spherical polar coordinates.  The transformation is:
\ba
x_1=&x_2+R \sin \theta\nonumber\\
y_1=&y_2 \hbox to .5in{}\\
z_1=&z_2+R \cos\theta\nonumber
\ea
where both $R$ and $\theta$ are constants.  The new origin may be considered to be the point from which atoms are launched in the lab.  The kinetic energy becomes
\ba
K=\frac{1}{2}m_i\bigg(\dot x_2^2+\dot y_2^2+ \dot z_2^2 \hbox to 1in{}\nonumber\\
 +2 \omega ((x_2+R \sin \theta)\dot y_2 - y_2 \dot x_2)\hbox to .4in{} \nonumber\\
+\omega^2(x_2^2+y_2^2 + 2 x_2 R \sin \theta + R^2 \sin^2 \theta)\bigg) 
\ea
The terms linear in $R$ in this equation will be seen to be important; they arise from both centripetal and coriolis effects.

	{\it Tilting so $z$ is parallel to the local vertical\/.}  Let $\lambda$ be the local value of geographic latitude where the fountain is located.  To tilt the $z_2$ axis through this angle, the transformation is
\ba
x_2 = &x \sin \lambda + z \cos \lambda ;\nonumber\\
y_2=&y;\hbox to .8in{}\\
z_2=&-x \cos \lambda+ z \sin \lambda. \nonumber
\ea
The angle $\lambda$ is a constant; differentiating to find the velocities in the new system, the kinetic energy becomes:

\begin{figure}\label{fig1}
\includegraphics[width=3.5 truein]{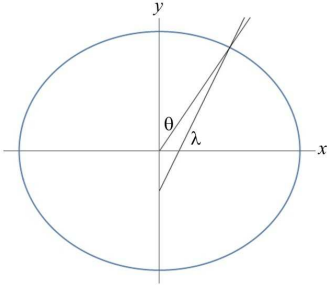}
\caption{The angles $\theta$ and $\lambda$ are illustrated in this figure.  $\theta$ is the angle measured downward from the north pole to the radial line from the origin to the position of the fountain.  $\lambda$ is the angle of the normal to the oblate surface that passes through the fountain position and can be measured from the equatorial plane.}
\end{figure}
\ba
K=\frac{1}{2}m_i\bigg(\dot x^2 +\dot y^2+\dot z^2+R^2 \omega^2 \sin^2\theta\hbox to 1in{}\nonumber\\
+2 \omega^2 R \sin \theta (x \sin \lambda + z \cos \lambda)\hbox to 1.5in{}\nonumber\\
+\omega^2((x \sin \lambda+z \cos \lambda)^2+ y^2)\hbox to 1.2in{}\\
+2\omega\big((x \sin \lambda+z \cos \lambda)\dot y\nonumber\hbox to 1in{}\\
+R \sin \theta \dot y-y(\dot x \sin \lambda + \dot z \cos \lambda)\big)\bigg)\,.\nonumber
\ea
The angles $\theta$ and $\lambda$ are illustrated in Figure 1.  For an ellipsoid with equatorial and polar radii $a$ and $b$, respectively, the relation between $\theta$ and $\lambda$ is
\be
\cos\lambda=\frac{b^2 \sin \theta}{\sqrt{b^4 (\sin \theta)^2+a^4 (\cos \theta)^2}}\,.
\ee

	{\it Tilting about the southerly direction\/}. Up to this point the axes have been tilted in a meridianal plane.  To account for the lack of azimuthal symmetry of the real gravitational field, we could perform one additional tilt about the $x$ axis, by an angle that we could call $\psi$, so that the final $z-$axis will be in the true local, geographically vertical direction.  We shall not actually do this in this paper because it complicates the algebra but it could be done in a straightforward manner. 

	{\it Gravitational Potential\/.}  We shall assume the gravitational potential is azimuthally symmetric.  This assumption allows one to consider terms in the gravitational potential that contribute to the overall oblateness, such as the quadrupole moment coefficient $J_2$.  The gravitational potential is now expanded about the new origin, and with this assumption there will be no gravitational field gradient in the $y-$ direction, although we include it temporarily:
\be
V(x,y,z)=V_0+\parderiv{V}{x}x+\parderiv{V}{y}y+\parderiv{V}{z}z\,.
\ee
The partial derivatives are evaluated at the origin. Only linear terms will be kept since the size of the fountain is very small compared to the radius $R$.

We now can write the Lagrangian function for an atom and study the equations of motion.  We allow, for the time being, that the gravitational mass may differ from the inertial mass so the Lagrangian is
\ba\label{lagrangian}
L=K-m_g V =\nonumber\hbox to 2in{}\\
\frac{1}{2}m_i\bigg(\dot x^2 +\dot y^2+\dot z^2+R^2 \omega^2 \sin^2\theta\hbox to 1.0in{}\nonumber\\                                                                          +2 \omega^2 R \sin \theta (x \sin \lambda + z \cos \lambda)\hbox to 1.in{}\nonumber\\
+\omega^2((x \sin \lambda+z \cos \lambda)^2+ y^2)\hbox to 1in{}\nonumber\\
+2\omega\big((x \sin \lambda+z \cos \lambda)\dot y+R \sin \theta \dot y\hbox to .5in{}\nonumber\\
-y(\dot x \sin \lambda + \dot z \cos \lambda)\big)\bigg)\hbox to .75in{}\nonumber\\
-m_g\big(V_0+\parderiv{V}{x}x+\parderiv{V}{y}y+\parderiv{V}{z}z \big).\hbox to .5in{}
\ea                                                     
 
\section{Equations of motion of an atom}
	The equations of motion can now readily be derived, using Eq. (\ref{lagrangian}) with the help of the Euler-Lagrange equations.  For the motion in the $y$-direction (eastward on earth's surface), we obtain
\ba
\frac{d}{dt}\bigg(m_i \dot y+ m_i \omega(x\sin \lambda + z \cos \lambda)+m_i \omega R \sin \theta\bigg)\nonumber\\
=m_i \omega^2 y- m_i \omega (\dot x \sin \lambda + \dot z \cos \lambda)-m_g \parderiv{V}{y}\,,\hbox to .25in{}
\ea
or, since $R$ and $\theta$ are constants, this simplifies to
\be
m_i \ddot y =m_i \omega^2 y-2 m_i \omega(\dot x \sin \lambda + \dot z \cos \lambda)-m_g \parderiv{V}{y}\,.\hbox to .25in{}
\ee
An atom at rest at the origin should suffer no acceleration in the $y-$direction, and this is satisfied if
\be
\parderiv{V}{y}=0\,,
\ee
as we have assumed.  The term $m_i\omega^2 y $ will be extremely small and we shall neglect it.  Then the resulting equation of motion for $y$ can be integrated:
\be
\dot y=-2 \omega (x \sin \lambda + z \cos \lambda)\,.
\ee
This result expresses conservation of angular momentum, coming through the Coriolis force in the rotating frame.  

Next consider the equation of motion in the $x$-direction.  The Euler-Lagrange equation for $x$ is:
\ba
\frac{d}{dt}(m_i \dot x-m_i \omega y \sin \lambda  )=m_i \omega^2 \sin \lambda(x \sin \lambda +\nonumber\\
 z \cos \lambda)+m_i \omega^2 R \sin\lambda \cos \theta\nonumber\hbox to .5in{}\\
\hbox to .5in{}	+m_i \omega \dot y\sin \lambda  -m_g \parderiv{V}{x}\,.\hbox to .5in{}
\ea
An atom at rest at the origin will suffer no acceleration in the $x$-direction if
\be
m_i \omega^2 R \sin\lambda \cos \theta-m_g\parderiv{V}{x}=0\,.
\ee
This condition basically expresses the fact that the vertical direction is at right angles to the $x$-direction, which is geographically South.  The remaining terms give
\be
\ddot x = -3 \omega^2 \sin \lambda (x \sin \lambda + z \cos \lambda)\,.
\ee
These terms are negligible.  We shall return later to estimate their contributions to the frequency shift.  If atoms are launched vertically upwards, there will be no appreciable motion in the Southerly direction.  In the following derivations, we shall therefore assume that $x=0,\ \dot x =0$.  

	The Euler-Lagrange equation for $z$ is then
\ba
m_i \ddot z = m_i \omega^2 \cos \lambda (x \sin \lambda + z\cos \lambda)+ m_i \dot y\omega^2 \cos \lambda\nonumber\\
+m_i \omega^2 R \cos \lambda \sin\theta -m_g \parderiv{V}{z}\,.
\ea    
For an atom at the origin with no velocity in the $y-$direction, the force is
\be
m_i \ddot z = m_i \omega ^2 R \cos \lambda \sin \theta -m_g \parderiv{V}{z}\,.
\ee
We shall abbreviate this by introducing the local value of the ``acceleration of gravity" at the lab, calling it $g$:
\be
\ddot z = -g = -\frac{m_g}{m_i}\parderiv{V}{z}+\omega^2 R \cos \lambda \sin \theta\,.
\ee
The local centripetal acceleration is included in the value of $g$; this contribution reduces the downward acceleration due to the gravitational potential gradient.  The acceleration $g$ includes both effects.  Motion in the vertical direction is just what one would expect. 

\section{Solutions of equations of motion}

	We may now state how an atom which is launched vertically upwards moves in the local, tilted, rotating frame.  If the launch velocity is $v_0$, then starting at $t=0$ we have
\be
\dot z = v_0-g t;
\ee
\be
z=v_0 t-\frac{1}{2} g t^2.
\ee
We assume the atoms rise to height $h$ and fall back down to their starting points.  The total time of flight we denote by $t_f$ and
\be
t_f = \frac{2 v_0}{g};\quad v_0 = \sqrt{2 g h}\,.
\ee
In the $y$-direction we have
\be
\dot y=-2z \omega \cos \lambda  = -2\omega \cos \lambda (v_0 t-\frac{1}{2} g t^2);
\ee
\be
y=-2 \omega \cos \lambda \big(\frac{1}{2}v_0 t^2 - \frac{1}{6} g t^3\big)\,.
\ee
We shall continue to assume there is no significant motion in the $x-$direction.

\section{Fractional frequency shifts}

	The instantaneous fractional frequency shift due to motion is just $-v^2/(2 c^2)$ and this can be obtained from the kinetic energy by multiplying by $-(1/m_i c^2)$.  To obtain the average fractional frequency shift we average over the time using Eq. (\ref{eq6}).  The shift will be obtained relative to a reference oscillator that is at rest at the origin.  For such an oscillator the only term in the kinetic energy that contributes is $\omega^2 R^2 \sin^2\theta/2$, so we drop this term from the time average of the shift of the launched atom.  Then the only terms that contribute significantly arise from the motion in the $z$-direction and from the terms linear in $R$ in the kinetic energy.  All other terms are smaller by factors of order $h/R$ and can be shown to be negligible.  Thus, the only terms in the kinetic energy that contribute will be
\be
\frac{1}{2}\big(2 \omega^2 R \sin \theta \cos \lambda z+2\omega R \sin \theta \dot y + \dot z^2\big)\,.
\ee
We then have
\be 
-\frac{1}{c^2}< \omega^2 R \sin \theta \cos \lambda z>=-\frac{2\omega^2 \sin \theta \cos \lambda} {3 c^2} h\,;
\ee
\be 
-\frac{1}{c^2}<\omega R \sin \theta \dot y>=\frac{4\omega^2 \sin \theta \cos \lambda} {3 c^2} h\,;
\ee
\be
<-\frac{\dot z^2}{2 c^2}>=-\frac{1}{3}\frac{g h}{c^2}\,.
\ee
The net fractional second-order Doppler contribution to the fractional frequency shift is thrn:
\be
\frac{\Delta f}{f}\bigg|_{Doppler}=-\frac{1}{3}\frac{g h}{c^2}+\frac{2 \omega^2 R \sin\theta \cos \lambda}{3 c^2} h\,.
\ee
Relative to a reference oscillator at the origin, the gravitational part of the frequency shift will be
\be
\frac{1}{c^2}<\parderiv{V}{z}z>=\frac{2}{3c^2}\parderiv{V}{z} h\,.
\ee
Since 
\be
\parderiv{V}{z}=g+\omega^2 R \sin \theta \cos \lambda\,,
\ee
the total gravitational fractional frequency shift is
\be
\frac{\Delta f}{f}\bigg|_{grav}=\frac{2}{3}\frac{g h}{c^2}+\frac{2}{3}\frac{\omega^2 R \sin\theta\cos\lambda}{c^2}h\,.
\ee
\section{Final result}

	We have assumed equality of inertial and gravitational mass.  Adding the contributions from second-order Doppler and true gravitational effects,
the net fractional frequency shift is
\be\label{finalresult}
\frac{\Delta f}{f}\bigg|_{total}=\frac{1}{3}\frac{g h}{c^2}+\frac{4}{3}\frac{\omega^2 R \sin\theta\cos\lambda}{c^2}h\,.
\ee

For example, at $\lambda=40^{\circ}$ N latitude, with $a=6378136.3$ m and $b=6356751.5$ m, $\theta=50.1894^{\circ}$, $g=9.796022\ {\rm m}/{\rm s}^2$, $R=6369344.1$ m.  For a height $h=1$ m the two terms in Eq. (\ref{finalresult}) are
respectively $3.63 \times 10^{-17}$ and $2.95 \times 10^{-19}$.
The latter correction will ordinarily be negligible for atomic fountain clocks.
\section{Appendix}
	
	We return to the kinetic energy to verify that remaining terms not having a factor of $R$ are negligible.  For example, one such term is
\be
<\frac{1}{2c^2}\omega^2 \cos^2 \lambda (v_0 t-\frac{1}{2} g t^2)^2>=\frac{8 \omega^2 \cos^2 \lambda h^2}{30 c^2}\,.
\ee
This terms and others like it are smaller than the terms we have retained by
factors of $h/R \approx 10^{-7}$.  We therefore have neglected them.  One can see by dimensional analysis that they are all of this order.

\end{document}